\documentstyle[preprint,aps]{revtex}
\begin{document}
\draft
\title{Comment on "Role of heavy meson exchange in near
threshold $N\! N \rightarrow d\pi$"$^*$ }
\author{J. A. Niskanen}
\address{ Department of Physics,
 P. O. Box 9, FIN--00014 University of Helsinki, Finland}

\maketitle
\begin{abstract}
In a recent paper by C. J. Horowitz (Phys. Rev. C {\bf 48},
2920 (1993)) a heavy meson exchange is incorporated
into threshold $N\! N \rightarrow d\pi$ to enhance the
grossly underestimated cross section. However, that
calculation uses an unjustified assumption on
the initial and final momenta, which causes an overestimate
of this effect by a factor of 3--4. I point out that the
inclusion of the $\Delta(1232)$ isobar increases the cross
section significantly even at threshold.
\end{abstract}

A recent paper \cite{horo} proposes that
heavy meson exchange (HME)
involving a nucleon-antinucleon pair may be important
in threshold pion
production in the reaction $np \to d \pi^0$ and $pp \to d
\pi^+$ (here generically included in the first reaction).
This mechanism contributes to the two-nucleon axial charge
\cite{lee,hmg}, and so far has
been the only way to explain the surprisingly large  $pp
\to pp \pi^\circ$ cross section at threshold \cite{meyer}.
The importance of this effect in this reaction is partly due
to the absence of charge-exchange pion $s$-wave rescattering,
dominant in the present $np \to d \pi^0$.
A motivation for the inclusion of the HME mechanism to the
deuteron reaction in Ref. \cite{horo} is the
stated underestimation of the cross section
\cite{hutch} by theory almost by a factor of two.
This addition  to the conventional
one-nucleon axial charge and $s$-wave pion
rescattering doubles the $s$-wave cross section bringing the
calculated results close to the data.

The aim of this Comment is to criticize an approximation
used in Ref. \cite{horo}, which  exaggerates the
HME effect in this reaction. The $\sigma$ meson exchange
leads to the operator
\begin{equation}
{\cal M}_{fi} \propto \frac{{\bf \sigma}_i\cdot
({\bf p' + p})} {2 M}\,\frac{1}{M}\,
\frac{g_\sigma^2}{m_\sigma^2 +{\bf k}^2}\,\tau_{i0}
\end{equation}
for each nucleon $i$.
Except for the momentum transfer dependent $\sigma$ propagator
this is similar to the Galilean invariance (axial charge)
part of the direct production operator. Exchange of the
other important
$\omega$ meson has an additional part $\propto {\bf \sigma_1
\times \sigma_2}$, which changes the spin and does not
contribute to $s$-wave production here. Eventually this
operator leads to radial integrals (with an opposite sign
convention from Ref. \cite{horo})
\begin{eqnarray}
J_\sigma & = & {g_\sigma^2\over 4\pi} \int_0^\infty \left[ \,
 (\frac{d}{dr} -\frac{1}{r}) v(r)
j_0 ({qr\over 2})\,            \nonumber
({e^{-m_\sigma r}\over 2Mr}) \, u_{11}(r)  \right. \\
&  - & \left. v(r)  \, j_0 ({qr\over 2})\,
({e^{-m_\sigma r}\over 2Mr}) \, (\frac{d}{dr} + \frac{1}{r} )
 u_{11}(r) \right]\, dr
\end{eqnarray}
for the deuteron $S$-state and
\begin{eqnarray}
J_\sigma & = & {g_\sigma^2\over 4\pi} \int_0^\infty \left[
(\frac{d}{dr} + \frac{2}{r}) w(r) \,
j_0 ({qr\over 2})\,           \nonumber
({e^{-m_\sigma r}\over 2Mr}) \, u_{11}^*(r) \right. \\
& - &\left. w(r) \, j_0 ({qr\over 2})\,
({e^{-m_\sigma r}\over 2Mr}) \,
(\frac{d}{dr} - \frac{2}{r} ) u_{11}^*(r) \right]\, dr.
\end{eqnarray}
for the $D$-state, with the derivatives acting only on the
nearest wave function. Similar equations are valid also for
$\omega$ exchange.

Following Koltun and Reitan \cite{kr}, Eqs. (25--30) of
Ref. \cite{horo} seem to
replace the momentum operator $({\bf p' +  p})$ operating
on both the initial and the final state wave functions with
$2{\bf p}$, because the pion momentum does not significantly
affect $s$-wave production at threshold. Although valid for the
direct production part, this is no more allowed in the presence
of the momentum
transferring HME potential, which does not commute with the
momentum operator. This assumption only picks (double) the
latter terms in the above equations. Elimination of the
derivative in the final state by integration by parts does
not help, since a derivative of the potential emerges. The first
line in Table \ref{table} shows these integrals using $2{\bf p}$
for the momentum operator and agrees well with Ref. \cite{horo}.
Correcting this approximation essentially halves the contribution
from the deuteron $S$-state, since there the final state
momentum contribution is small. However, for the $D$-state this
is significant and in destructive interference with the main part,
as can be seen from Table \ref{table}.  Overall the HME
contribution to the amplitude is decreased by a factor of
4 and cannot account for the missing cross section. Instead
of an increase of the conventional $\alpha$ by 86 $\mu$b reported
in Ref. \cite{horo}, the change is now 18 $\mu$b. In these
calculations the Bonn potential A(R) $\sigma$ and $\omega$
couplings and form factors are used with Reid soft core wave
functions as in Ref. \cite{horo}.

The same approximation
is used also for $pp \to pp \pi^\circ$ in Ref. \cite{hmg}.
Therefore, as a check, it was established
that in $pp \rightarrow pp\pi^0$ the $\sigma$ and $\omega$
exchanges alone give a good description of the low energy data
\cite{meyer}. With the more precise treatment of the final
state momentum, however, the $\sigma$ contribution decreases
to nearly a half, while the $\omega$ effect is enhanced enough
to compensate this loss. It may be noted that
the $\sigma$ and $\omega$  mesons were by far the most
important in Ref. \cite{hmg}.

However, the present reaction is more involved
than is apparent from the above discussion. The
threshold description of Koltun and Reitan \cite{kr},
employed in Ref. \cite{horo} with modern two-nucleon potentials
and deuteron wave functions does not consider the role of an
explicit virtual $\Delta$(1232) isobar excitation in producing
pions. This dominates $p$-wave production. Although the
centrifugal barrier in the $P$-wave baryon states suppresses
the $\Delta$ components to some extent, it can be seen
from Fig.\ 1 that even for threshold $s$-wave pions the isobar
effect is by no means negligible and its inclusion triples
the cross section. Therefore, in a more complete model the
threshold cross section may be actually {\it overestimated}
even without HME. By far most of this increase in $s$-wave pion
production comes from the normal elementary ($p$-wave) emission
of the pion from the $\Delta$ followed by $s$-wave rescattering
from the second nucleon.  Adding HME  slightly increases the
overestimation as shown in Fig.\ 1.

 The present calculation treats the $\Delta$ isobar on
the same basis as the nucleons by a system of
coupled Schr\"odinger equations for a modified Reid soft core
potential with $N\!\Delta$
admixtures in the initial $N\! N$ state \cite{ppdpi}.
This causes also some short range changes in the $N\! N$ wave
function reflected in HME as seen in the third line of
Table \ref{table}. HME is included only in the nucleon sector.
Energy dependence is allowed for $s$-wave pion rescattering
to fit on-shell $\pi N$ scattering \cite{impact},
but except for the virtual $N\!\Delta$ admixtures  the model
reduces to the formalism of Koltun and Reitan at threshold.
(A monopole form factor with $\Lambda = 700$~MeV is also
included to account for off-shell rescattering.)

Another aspect for caution in adding new mechanisms to threshold
amplitudes is the changes caused in observables at higher energies,
where there are much more data to basically fix the amplitudes.
 The analyzing power $A_y$ between 500 and 600 MeV is particularly
sensitive to the $s$-wave pion amplitude. In this region
 the coupled-channels model used above to produce the solid
curves has been particularly successful.  The use of a smaller
$s$-wave amplitude to fit the threshold cross section would
produce too high an analyzing power, whereas a larger one would
yield too deep a minimum in it. Again it is lucky that the
HME effect is small in this reaction. Further, since the low energy
analyzing power data \cite{korkmaz} can be easily fitted by simply
scaling the $s$-wave amplitude with the factor
$\sqrt{\sigma({\rm th})/\sigma{\rm (exp)}}$, which
compensates for the overestimation of the $s$-wave amplitude,
one may conclude that apparently the $p$-wave amplitude is under
control also close to threshold.

The overestimation of the threshold cross section by the
full model, which agrees well with the data in the $\Delta$
region, poses a slight problem of detail indicating that
either the energy dependence of off-shell pion rescattering
is not properly incorporated or some physical mechanism
is still missing in the present models of pion production.
However, HME does not appear important here, whereas a
significant contribution from the $\Delta$ is likely to
survive improvements of the model.

\begin{figure}
\caption{ Low-energy  $pp \to d \pi^+$  cross section
divided by $\eta = q_\pi/m_\pi$.  The solid curves
show the starting point before the addition of HME with
the $\Delta$ included in all partial waves (the lower
one is the $s$-wave contribution), while
the dashed curve is the  $s$-wave contribution
without the  $\Delta$. The dotted and dash-dot curves
have also the HME added to these calculations of the $s$-wave.
The data are from Ref.\ \protect\cite{hutch}.}
\end{figure}

\begin{table}
\caption{Integrals of Eqs. (2,3) (in fm$^{-1/2}$) for $\sigma$
and $\omega$ exchanges and $S$ and $D$ final states for $\eta
= q_\pi/m_\pi = 0.1424$. The
total has also a factor $1/\protect\sqrt{2}$ multiplying the
$D$ state as required by angular momentum algebra
\protect\cite{kr}.}
\begin{tabular}{cccccc}
Model & $\sigma ,\, S$ & $\sigma ,\, D$ & $\omega ,\, S$ &
$\omega ,\, D$ & Total \\ \hline
$2{\bf p}$ & -0.0700 & -0.0519 & -0.0004 & -0.0002 & -0.1223 \\
${\bf p' +  p}$ & -0.0284 & -0.0202 & 0.0162 & 0.0119 &
-0.0287 \\
$N\Delta$ & -0.0260 & -0.0181 & 0.0059 & 0.0037 & -0.0373 \\
\end{tabular}
\label{table}
\end{table}

\end{document}